de' Broglie's paradox, relativistic Doppler effect, and the derivation of mass-energy relation in special relativity

Guang-jiong Ni[a)]

Department of Physics, Fudan University, Shanghai 200433, China
Department of Physics, Portland State University, Portland, OR97207, U.S.A.

Abstract. Three related topics are discussed to show the "simplicity, harmony and beauty" of the theory of special relativity (SR): (1) How can de' Broglie discover his famous relation between the particle's momentum and a wave length--- a paradox stemming from the time dilatation effect of SR. (2) The relativistic Doppler effect, especially the distinction between the transverse Doppler effect and the time dilatation effect. (3) Three particular examples are examined to derive the general mass-energy relation in SR by induction method.

## I. INTRODUCTION

The theory of special relativity (SR) established by Einstein in 1905 and the theory of quantum mechanics (QM) are two great pillars of the whole framework of modern physics. The purpose of this paper is to emphasize SR being the direct promoter of QM (section II) and discuss three elegant examples (sections V, VI and the Appendix) to show the power of the principle of relativity in SR as well as the essence of induction method. Section III is a simple review of relativistic Doppler effect because it is intimately related to sections V, VI and there is some confusion in textbooks about the distinction between the time dilatation effect and the transverse Doppler effect (section IV). The final section VII contains a summary and discussion. Although there is essentially nothing new in this paper, which might still be useful as a reference in teaching SR.

## II. DE'BROGLIE'S PARADOX AND ITS SOLUTION

As is well known, QM, especially its form invented by Schrödinger, was initiated from the seminal work by de'Broglie in 1923. An equation named after him is taught in every course of QM:
$$\lambda = h/p \qquad (2.1)$$
where $\lambda$ is the wave length of "de'Broglie wave" associated with a particle having momentum p and h is the Planck constant. However, one thing needs to be stressed is how could de'Broglie discover Eq. (2.1)? To relate a wave to a particle was an absurd idea at that time. How could he reach this idea?
The clue lies in a paradox raised by him. First, a famous relation was already proposed by Einstein in 1905 to explain the photo-electric effect:
$$E = h f \qquad (2.2)$$

Where E is the energy of a "photon" with f being the frequency of associated electromagnetic wave. Eq. (2.2) also seemed absurd in 1905 because in classical electrodynamics the energy of wave should be directly proportional to its amplitude square rather than frequency. As Einstein's relation (2.2) had been tested in numerous experiments, de'Broglie boldly assumed that it might be also valid for a particle like electron. It means that there might be some vibration associated with an electron and its frequency f is proportional to the electron's energy E:

$$E = hf = \frac{m_o c^2}{\sqrt{1 - v^2/c^2}} \qquad (2.3)$$

where $m_o$ (v) is the rest mass (velocity) of electron while c is the speed of light. When v = 0, (2.3) reads:

$$hf_o = m_o c^2 \qquad (2.4)$$

So

$$f = \frac{f_o}{\sqrt{1 - v^2/c^2}} \qquad (2.5)$$

As a next conjecture, he might think that the vibration is belong to an "inner clock" moving with the electron as often discussed in the theory of SR. However, as is well known in SR, if the clock has a frequency $f_o$ in the rest frame of electron, its frequency would slow down to $f_1$ when it is moving at a velocity v:

$$f_1 = f_o \sqrt{1 - v^2/c^2} \qquad (2.6)$$

Evidently, the so-called time dilatation effect (2.6) is just opposite to Eq. (2.5). Hence we may name the contradiction between them a " de'Broglie's paradox".

There are a number of paradoxes in physics, aiming at pushing the contradiction of theory into such an acute situation so that one can find what was wrong in the old concept and/or what new ingredient was missing in the old theory. What de'Broglie eventually realized is that f and $f_1$ are two different things: While $f_1$ is the frequency of an "inner clock" moving with the electron and an observer( staying in the laboratory (L) frame) must arrange his apparatus moving at a same velocity before $f_1$ can be measured, the f should be another frequency of a "wave' associated with the particle. To measure f of a wave, the observer keeps measuring its vibration frequency at a fixed point in the L frame. As the conditions of measurement are different, the difference between f and $f_1$ should be understandable. There is no contradiction at all.

As a third step, de'Broglie resorted to the whole theory of SR in two inertial frames, the rest frame of electron and the L frame. With the bold hypothesis (2.1) as the wavelength of his "wave", he was capable of making the whole theory self-consistent.Actually, one more assumption ( called by de'Broglie as the "phase harmony law" and regarded by himself as the most fundamental contribution all his life[1]) was used, saying that the phase of "wave" is an invariant of the Lorentz transformation.

In some sense, de'Broglie discovered the half of quantum theory, Eq. (2.1), from another half, Eq.(2.2), by means of the whole theory of SR. His genius thinking really won a great success in the history of physics.

However, the de'Broglie wave (later developed into "wave function" in general by Schrödinger) is not observable. How can we measure f and $f_1$ in real experiments? See next section.

## III. TIME DILATATION AND RELATIVISTIC DOPPLER EFFECT

Consider that a moving source with velocity v (along the x axis of laboratory (L) frame) emits light with frequency $f_0$ in its rest (R) frame, implying an intrinsic period $T_o = 1/f_o$. However, in the L frame, two successive wave crests are emitted at a short time interval

$$\Delta t = t_2 - t_1 = \gamma T_o = \gamma/f_o \qquad (3.1)$$

where
$$\gamma = 1/\sqrt{1-\beta^2}, \qquad \beta = \frac{v}{c} \qquad (3.2)$$

Eq.(3.1) is just (2.6), showing the time dilatation effect of SR. Moreover, for an observer located at a fixed point Q on the x axis of L frame, the time interval between arrivals of these two wave crests should be calculated as [2]:

$$T = t_2 + r_2/c - (t_1 + r_1/c) = \gamma/f_o - (r_2 - r_1)/c = (1 - \frac{v}{c}\cos\theta)\gamma/f_o \qquad (3.3)$$

since $x_2 - x_1 \ll r_1, r_2 - r_1 \approx (x_2 - x_1)\cos\theta = v(t_2 - t_1)\cos\theta$, where $\theta$ is the angle between the light and the x axis of L frame. Hence the light frequency measured by the observer is [2]:

$$f = \frac{1}{T} = \frac{f_o\sqrt{1-\beta^2}}{1-\beta\cos\theta} \qquad (3.4)$$

Notice that, however, if we consider the angle $\theta'$ between the light and the x' axis (which is parallel to x axis with distance between x and x' axes being much larger than $cT_o$) in the rest frame of source, a relation between $\cos\theta$ and $\cos\theta'$ can be deduced by the Lorentz transformation as [2]:

$$\cos\theta = \frac{\cos\theta' + \beta}{1+\beta\cos\theta'}, \qquad \cos\theta' = \frac{\cos\theta - \beta}{1-\beta\cos\theta} \qquad (3.5)$$

Substituting (3.5) to (3.4) leads to:

$$f = \frac{f_o(1+\beta\cos\theta')}{\sqrt{1-\beta^2}} \qquad (3.6)$$

It is interesting to see the same frequency f being expressed in two different forms (3.4) and (3.6). Four cases will be discussed:

(a) Nonrelativistic case: $\beta = v/c \ll 1, \beta^2 \approx 0, \cos\theta \approx \cos\theta' + \beta$,

$$f = f_o(1 + \beta\cos\theta') \approx f_o(1 + \beta\cos\theta) \qquad (3.7)$$

This is the classical Doppler effect at the first order of $\beta$, showing frequency change due to a relative motion between the source and receiver.

(b) Relarivistic case, $\beta \to 1$: Consider $\cos\theta = 0$ in (3.4) first. We find:

$$f = f_o\sqrt{1-\beta^2} \qquad (3.8)$$

which is just (2.6) showing the time dilatation effect of SR. Obviously, the condition of measurement $\theta = 90°$ means that the observer(Q) must keep moving at the same velocity v with the light source so that he can receive the light perpendicular to his x axis. Note that the relativistic effect always emerges as a second order ($\beta^2$) effect.

(c) Relativistic case, $\beta \to 1$ again, but $\cos\theta' = 0$ in (3.6): Instead of (3.8), we find:

$$f = \frac{f_o}{\sqrt{1-\beta^2}} \qquad (3.9)$$

which is called the transverse Doppler effect. Eq. (3.9) coincides precisely with (2.5), in this case the light frequency reflects the frequency of de'Broglie wave.

Both cases (b) and (c) will be discussed in subsequent sections.

(d) Relativistic longitudinal Doppler effect: If $\theta = \theta' = 0$, either (3.4) or (3.6) gives:

$$f = f_o\sqrt{\frac{1+\beta}{1-\beta}} \qquad (3.10)$$

which approaches Eq. (3.7) when $\beta \to 0$. Eq.(3.10) could be understood as the first order (classical) Doppler effect (3.7) modified by the second order transverse Doppler effect (3.9), (rather than by the time dilatation effect (3.8)).

## IV. EXPERIMENTAL VERIFICATION OF RELATIVISTIC DOPPLER EFFECT AND GRATITATIONAL REDSHIFT

The relativistic Doppler effect was proposed by Einstein in 1907 and first confirmed by the experiment of Ives and Stilwell in 1938[3]( see also [2,4]). Then two experiments were conducted by MacArthur et al. in 1986[5] and by McGowan et al. in 1993[6] respectively. Both of them used the mutual interaction between laser beam and atom beam and so achieved high accuracy in the verification of Eq. (3.4). While in MacArthur's experiment with $\beta = 0.84$ the accuracy is $2.4 \times 10^{-4}$, which is improved to $2.3 \times 10^{-6}$ in McGowan's experiment with $\beta = 0.0036$. Notice that the angle $\theta$ in both papers [5] and [6] is that between the laser beam and atom beam directions, so their $\theta$ is precisely that in Eq. (3.4), an angle that can only be measured in the laboratory. In fact, the angle $\theta'$ in Eq. (3.6), which means the direction of light beam in the rest frame of light source, is very difficult (even impossible) to measure in their experiments.

However, there is another kind of experiment, first realized by Hay et al. in 1960[7] and by kündig in 1963[8] (see discussion by Sherwin[9]). They predicted that if using a rotating system with angular velocity $\omega$ and putting the (light or $\gamma$) source (S) near the center, the absorber(A) at the rim of rotator with radius $R_A$ will detect a fractional energy change:

$$(E_A - E_S)/E_S = (1 - \beta^2)^{1/2} - 1 \approx -\frac{1}{2}\beta^2 = -R_A^2 \omega^2/2c^2 \qquad (4.1)$$

where $E_A$ and $E_S$ are the characteristic energies of the absorber and the source.

In Eq. (4.1), there are two different modes to explain the same effect: one is resorting to SR----the absorber is moving at a speed $v = \omega R_A$, $(\beta = v/c)$ in the laboratory. So its "inner clock" (or atom) is vibrating at a slower frequency as shown by Eq. (2.6). Notice also that the photon is received by the absorber along the radius direction perpendicular to the direction of absorber's motion, i.e., $\theta = 90°$ in Eq. (3.4). Hence the relativistic Doppler effect in Eq. (4.1) (which was often called the "transverse Doppler effect" in some literatures and text books, but we prefer to refer it being the time dilatation effect) exhibits itself as a decreasing of frequency. Another explanation of Eq. (4.1) is resorting to the theory of general relativity (GR): the frame K attached to the rotating absorber is an accelerating one, the principle of equivalence in GR interpret the centrifugal force acting on the absorber as a gravitational force with the potential

$$\Phi = -\frac{1}{2} R_A^2 \omega^2 \qquad (4.2)$$

Thus the observer in K will assert that his clock is slowed down by the gravitational potential, i.e., the characteristic frequency $f_A$ of incident light measured in K frame is lower than $f_S$ associated with the source where the gravitational potential is higher than that at absorber shown by Eq. (4.2):

$$f_A = f_S (1 + 2\Phi/c^2)^{1/2} \approx f_S (1 + \Phi/c^2) \qquad (4.3)$$

So Eq. (4.1) is proved theoretically and then verified experimentally up to an accuracy of 1.1% [8].

## V. DERIVATION OF MASS-ENERGY RELATION VIA A RESONANT ABSORPTION PROCESS

In his second paper for establishing SR in 1905, Einstein proposed the greatest equation changing the world:

$$E = mc^2 \qquad (5.1)$$

To derive it, Einstein had resorted to an ideal experiment, considering the interaction between electromagnetic wave with matter.(see also [2],p.16). However, many text books on SR often devote a great effort to discuss the process of elastic (or inelastic) collision between two particles for deriving (5.1) and the velocity dependence of m:

$$m = m_o \left/ \sqrt{1 - v^2/c^2} \right. \qquad (5.2)$$

It seems to us that once the derivation is not simple enough, it might obscure to some extent the main essence of the induction method, which is needed to establish a law like (5.1) with (5.2). Let us try to find a simple way of presentation which may be nearer to the original idea of Einstein.

First of all, we should note that the establishment of (5.1) with (5.2) is essentially a conjecture, i.e., a guess work of induction method---- we guess first a possible relation between the rest energy $E_o$ and the rest mass $m_o$ of a particle as:

$$E_o = m_o c^2$$

Then if it is moving at a velocity v, its kinetic energy $\frac{1}{2} m_o v^2$ should be viewed as the increment of E:

$$E = E_o + \Delta E = m_o c^2 + \frac{1}{2} m_o v^2 + ...$$

where .... means the possible corrections at high speed. To Einstein, Eqs. (5.1) and (5.2) could be guessed quite naturally. The next crucial step is to find an example for checking them rigorously.

Let us following Einstein, just consider a simple process of a rest particle (with rest mass $m_o$) absorbing resonantly a photon (with frequency f) because we already know that a photon has energy hf and momentum hf/c in the laboratory(S) frame. We will assume Eqs. (5.1) and (5.2) to see if the conservation laws hold in both S frame and S' frame in which the particle after absorbing the photon is at rest with rest mass $m_o'$.

Denoting the velocity of particle after absorbing the photon in the S frame by v, and the frequency of photon in the S' frame by f', we write down the momentum and energy conservation laws in the S frame as:

$$\frac{hf}{c} = \frac{m_o' v}{\sqrt{1 - v^2/c^2}} \qquad (5.3)$$

and

$$hf + m_o c^2 = \frac{m_o' c^2}{\sqrt{1 - v^2/c^2}} \qquad (5.4)$$

respectively. Similarly, in the S' frame, we have:

$$\frac{hf'}{c} = \frac{m_o v}{\sqrt{1 - v^2/c^2}} \qquad (5.5)$$

$$hf' + \frac{m_o c^2}{\sqrt{1 - v^2/c^2}} = m_o' c^2 \qquad (5.6)$$

Now we have four equations containing three unknown quantities v, $m_o'$ and $f'$ (with $m_o$ and f being given). If we can solve them consistently, we can then claim the validity of Eqs (5.1) with (5.2). To see this, we first solve from (5.3) and (5.5) that ($\beta = v/c$):

$$m_o' c^2 = \frac{hf}{\beta}\sqrt{1-\beta^2} \tag{5.7}$$

$$hf' = \frac{m_o c^2 \beta}{\sqrt{1-\beta^2}} \tag{5.8}$$

Next, substitute (5.7) to the right side of (5.4), yielding:

$$\beta = \frac{hf}{hf + m_o c^2} \tag{5.9}$$

Third, substituting (5.9) into the right side of (5.8), we find:

$$f' = f\frac{1-\beta}{\sqrt{1-\beta^2}} = f\sqrt{\frac{1-\beta}{1+\beta}} \tag{5.10}$$

which is exactly the longitudinal Doppler effect (3.10) . The final crucial step is substituting them altogether to Eq. (5.6) to see if it is valid. Then it turns to be just right--- both sides have the same value $m_o' c^2$ as shown in (5.7). All four equations match perfectly, the proof is finished. It means the validity of Eqs. (5.1) with (5.2) as well as the basic postulate of SR-----the principle of relativity does hold for two inertial frames. However, we should notice that being a law, (5.1) with (5.2) has to be verified eventually by experiments. The above derivation merely shows the essence of induction method ---- from particular ( the process of particle absorbing a photon) to general ( the validity of (5.1) and (5.2) in two inertial frames) . So the number of unknown variables must be less than that of equations. This is not a method of deduction-----from general (well-established theory) to particular------where the number of unknown variables should be equal to that of equations in one inertial frame so that we can get a particular solution unambiguously. In short, what we need to find from an induction method is a general law, not a particular solution. So the validity of law should be ensured by more and more particular experiments. To this purpose, we shall consider two more examples in the next section and the Appendix.

## VI.  ANNIHILATION RADIATION OF A POSITRONIUM

Similar to a hydrogenlike atom, a positronium is a bound state of electron and positron. It is unstable against decay to two photons with lifetime $10^{-10}$ sec. in its rest (R) frame:

$$e^+ + e^- \rightarrow \gamma_1 + \gamma_2 \tag{6.1}$$

Each one of these two photons, called the annihilation radiation, has energy $hf_o \sim m_e c^2 = 0.511\,\text{MeV}$ in the R frame in accordance with the energy conservation law:

$$m_o c^2 = 2hf_o \tag{6.2}$$

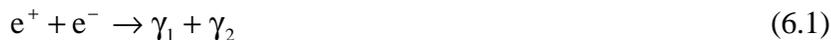

where $m_o$ is the rest mass of positronium ($m_o \leq 2m_e$). According to the momentum conservation law, if one photon is flying out along y axis, another one must be along – y axis-----a trivial equality providing no new information.

Let us assume that in the laboratory (L) frame, the positronium is moving at a velocity v along the x axis. Then the momentum conservation laws along x and y axes read:

$$2\frac{hf}{c}\cos\theta = \frac{m_o v}{\sqrt{1-\beta^2}} \qquad (6.3)$$

$$\frac{hf}{c}\sin\theta = \frac{hf_o}{c} \qquad (6.4)$$

where f is the frequency of photon $\gamma_1$ (or $\gamma_2$) in the L frame and $\theta$ is its angle with respect to x axis. Eq. (6.4) means that the y component of photon momentum is the same in R and L frames according to the Lorentz transformation. Meanwhile, the energy conservation law in the L frame reads:

$$2hf = \frac{m_o c^2}{\sqrt{1-\beta^2}} \qquad (6.5)$$

As in previous section, we have four equations (6.2)-(6.5) whereas only three unknown variables $f_o$, f and $\theta$ (with $m_o$ and v being given). While Eq. (6.2) simply gives the relation between $f_o$ and $m_o$, Eqs. (6.2) to (6.4) yield:

$$\cos\theta = \beta = v/c \qquad (6.6)$$

$$f = \frac{f_o}{\sqrt{1-\beta^2}} \qquad (6.7)$$

Thus we find all three variables $f_o$, f and $\theta$ (in terms of m and v). It is easy to check that Eq. (6.5) is satisfied perfectly. The proof is finished. Note that Eq. (6.7) is just a transverse Doppler effect as discussed in Eq. (3.9) because now the photon is emitted along a direction perpendicular to positronium's velocity v in its rest frame [$\theta' = 90°$ in Eq. (3.6)]. By the way, if an observer moves at a velocity along one of these two photons, he will see the longitudinal Doppler effect as shown by Eq. (3.10).

## VII. SUMMARY AND DISCUSSION

(a) SR is the direct promoter of QM not only because SR promoted the discovery of de'Broglie wave which led further to the invention of Schrödinger equation, but also because Einstein's idea directly promoted the invention of QM by Heisenberg along another route complementary to that of de'Broglie and Schrödinger, see [10].

(b) The Doppler effect is an effect of frequency change $\Delta f = f - f_o$ due to relative velocity v between the source and the observer. While $\Delta f/f_o$ is of the order v/c in classical physics, it is further modified to the second order $(v^2/c^2)$ in SR. Although various relativistic Doppler effects are stemming from the common essence of SR, we should be careful not to confuse the time dilatation effect (3.8) with the transverse Doppler effect (3.9) because their conditions in measurement are different. On the other hand, the gravitational redshift (GRS) is the frequency change due to the existence of gravitational field ,i.e., due to the relative acceleration between the source and the observer. Interesting enough, in experiments on the rotator, once GRS is taken into account, the second order Doppler effect in SR should not be doubly counted.

(c)  To derive the Einstein Eq. (5.1) with (5.2) by induction method, we present two examples in sections V and VI as well as another one in the Appendix. In all three cases, we must consider two inertial frames simultaneously so that the number of equations exceeds that of unknown variables. Thereby each case poses a test on the general validity of (5.1) with (5.2) as well as the principle of relativity.

(d)  The example in section V provides the strongest proof because once a photon---a totally relativistic object---- is coexisting with the particle ( before absorption), the principle of relativity—both momentum and energy conservation laws hold rigorously in two inertial frames ----- poses so stringent a constraint that other possibility except (5.1) and (5.2) is definitely excluded. The example in section VI seems a little weaker because the uniqueness of (5.1) with (5.2) is ensured by the validity of relativistic Doppler effect (6.7) and (3.10) (which must be derived from the Lorentz transformation.) of two photons after decay. The example in the Appendix seems even weaker because its unambiguous conclusion (5.1) with (5.2) must be ensured by the velocity addition law of Lorentz transformation (A.11) explicitly. Otherwise, one can always discuss the collision between two particles in the realm of classical mechanics (where mass and energy are two unrelated things) without violating the principle of relativity in the sense of Galilean transformation.

(e) Nevertheless, all topics discussed in this paper are capable of reflecting the "simplicity, harmony and beauty" of SR to some extent.

Acknowledgements: I thank S.Q. Chen, P.T. Leung, Y.S. Wang and Y.L. Zheng for relevant discussions.

## APPENDIX. INELASTIC COLLISION BETWEEN TWO IDENTICAL PARTICLES

As a further check of Eqs. (5.1) and (5.2), let us discuss a simplest example of two identical particles each with a rest mass $m_o$, colliding in a completely inelastic manner so that they form a compound particle with rest mass $M_o$. Again, we consider two inertial frames: In S frame, the first particle having a velocity u collides with the second one at rest, then after collision, the particle $M_o$ has a velocity v.  In another S' frame, the $M_o$ is at rest, so before collision, two particles have velocities v and – v.

Hence the momentum and energy conservation laws in the S frame read:

$$\frac{m_o u}{\sqrt{1-u^2/c^2}} = \frac{M_o v}{\sqrt{1-v^2/c^2}} \qquad (A.1)$$

$$\frac{m_o c^2}{\sqrt{1-u^2/c^2}} + m_o c^2 = \frac{M_o c^2}{\sqrt{1-v^2/c^2}} \qquad (A.2)$$

However, in the S' frame, only the energy conservation law is nontrivial;

$$\frac{2m_o c^2}{\sqrt{1-v^2/c^2}} = M_o c^2 \qquad (A.3)$$

As a whole, we have three equations but only two unknown variables $M_o$ and v (with $m_o$ and u being given). In SR, the following notations are proved to be very useful:

$$\beta_u = \frac{u}{c} = \tanh \varsigma_u, \quad \gamma_u = \frac{1}{\sqrt{1-\beta_u^2}} = \cosh \varsigma_u, \quad \beta_u \gamma_u = \sinh \varsigma_u \qquad (A.4)$$

where $\varsigma_u$ is called the rapidity of particle with velocity u. With similar notations for v, we recast Eqs. (A.1) - (A.3) into:

$$m_o \sinh \varsigma_u = M_o \sinh \varsigma_v \qquad (A.5)$$
$$m_o (\cosh \varsigma_u + 1) = M_o \cosh \varsigma_v \qquad (A.6)$$
$$2m_o \cosh \varsigma_v = M_o \qquad (A.7)$$

Now it is a pleasure to find consistently the solution of these three equations as:

$$\cosh \varsigma_v = \sqrt{\frac{1+\cosh \varsigma_u}{2}} = \cosh\left(\frac{\varsigma_u}{2}\right) \qquad (A.8)$$

$$M_o = 2m_o \cosh\left(\frac{\varsigma_u}{2}\right) \qquad (A.9)$$

Eqs.(A.8) and (A.9) are rigorous solution to the above relativistic dynamical problem in the whole range (0,c) of u, corresponding to $\varsigma_u$ ranging from 0 to infinity.

In the nonrelativistic limit (u<<c), we have v = u /2 and $M_o = 2m_o$ as expected.

However, we should note that the velocity v can be expressed in terms of u as [2]:

$$v/c = \tanh \varsigma_v = c/u\,[1-(1-u^2/c^2)^{1/2}] \qquad (A.10)$$

which is exactly that derived from the velocity addition law of Lorentz transformation :

$$u' = \frac{u-v}{1-uv/c^2} \qquad (A.11)$$

Indeed, setting u' = v in (A.11), we get (A.10) immediately.